\documentclass[preprint,showpacs,preprintnumbers,amsmath,amssymb]{revtex4}
\usepackage{epsfig}
\usepackage{amsmath}
\usepackage{amsfonts}
\usepackage{amssymb}
\usepackage{times}

\newcommand{\be}{\begin{equation}}
\newcommand{\ee}{\end{equation}}
\newcommand{\bea}{\begin{eqnarray}}
\newcommand{\eea}{\end{eqnarray}}
\newcommand{\ba}{\begin{array}}
\newcommand{\ea}{\end{array}}

\begin{document}

%
%
\title{Co-occurrence of resonant activation and noise-enhanced
stability in a model of cancer growth in the presence of immune response}

\author{Alessandro Fiasconaro}
\affiliation{Dipartimento di Fisica e Tecnologie Relative and CNISM,
Group of Interdisciplinary Physics\footnote
{http://gip.dft.unipa.it}, Universit\`a di Palermo, Viale
 delle Scienze, I-90128 Palermo, Italy}

\author{Anna \surname{Ochab--Marcinek}}
\affiliation{Marian~Smoluchowski Institute of Physics,
 Jagellonian University, Reymonta~4, 30--059~Krak\'ow, Poland}
\email{ochab@th.if.uj.edu.pl}

\author{Bernardo Spagnolo}
\affiliation{Dipartimento di Fisica e Tecnologie Relative and
CNISM, Group of Interdisciplinary Physics\footnote
{http://gip.dft.unipa.it}, Universit\`a di Palermo, Viale
 delle Scienze, I-90128 Palermo, Italy}

\author{Ewa \surname{Gudowska--Nowak}}
\affiliation{Marian~Smoluchowski Institute of Physics,
 Jagellonian University, Reymonta~4, 30--059~Krak\'ow, Poland}
\email{gudowska@th.if.uj.edu.pl}

\date{\today}

\begin{abstract}
We investigate a stochastic version of a simple enzymatic reaction
which follows the generic Michaelis-Menten kinetics. At
sufficiently high concentrations of reacting species, that
represent here populations of cells involved in cancerous
proliferation and cytotoxic response of the immune system, the
overall kinetics can be approximated by a one-dimensional
overdamped Langevin equation. The modulating activity of the
immune response is here modeled as a dichotomous random process of
the relative rate of neoplastic cell destruction. We discuss
physical aspects of environmental noises acting in such a system,
pointing out the possibility of coexistence of dynamical regimes
where noise-enhanced stability and resonant activation phenomena
can be observed together. We explain the underlying mechanisms by
analyzing the behavior of the variance of first passage times as a
function of the noise intensity.
\end{abstract}

\maketitle


\section{Introduction}


The fact which has gained recently considerable attention is that
randomly fluctuating systems can behave quite differently
from deterministic ones. On the one hand, noise may play a
destructive role in natural processes, leading to irregularities or
even completely random behavior. On the other hand, random
variations may, paradoxically, bring a system to a more ordered
state. Phenomena of this kind, e.g. noise-induced transitions,
stochastic resonance, noise-enhanced transport or noise-sustained
synchronization have been observed in diverse range of systems in
physics, chemistry, biology, and
medicine~\cite{Horsthemke_Lefever,Gammaitoni,Anishchenko}. It is
becoming apparent that fluctuations and noise are essential
ingredients of life processes. Inclusion of stochasticity in
mathematical models of biological and biochemical processes is thus
necessary for better understanding of mechanisms that govern the
biological systems.

The Michaelis-Menten mechanism for enzymatic catalysis is one of
the most important mechanisms in biochemistry. The most popular
nowadays application of the Michaelis-Menten kinetics is modelling
of intracellular biochemical regulation networks, where this sort
of kinetics is postulated as one of the standard types of
reactions, out of which complex models of cell behavior can be
constructed~\cite{Tyson_Novak-Tomita}. In a slightly modified
form, the Michaelis-Menten reaction scheme turned out to be of
practical use in biophysical modelling of radiation-induced damage
production and processing. In particular, this reaction scheme
 has been adapted for the purpose of studying kinetics of
double-strand breaks rejoining and formation of simple chromosome
exchange aberrations after DNA exposure to ionizing
radiation~\cite{Cucinotta}. Similar kinetics has been proposed in
saturable repair models of the evolution of radio-biological
damage~\cite{Kiefer-Sachs}.

Another field where the Michaelis-Menten kinetics finds a broad
application is population dynamics. Predator-prey models based on
that mechanism are very  often used in the analysis of population
dynamics of bacteria, plankton, plants, or
animals~\cite{Messier,Sala-Banik,Prigogine_Lefever} in various
ecosystems. The cell-mediated immune surveillance against cancer
is as well one of the effects which may be described in terms of a
"predator-prey" system and be approximated by a saturating,
enzymatic-like process whose time evolution equations are similar
to the standard Michaelis-Menten
kinetics~\cite{Zhivkov_Waniewski,Lefever_Horsthemke,Garay}. The
population of tumor cells plays here the role of "preys" whereas
the immune cells act as "predators". The activity of the predator
in a certain territory, or the activity of immune cells in tissue,
resemble the mode of action of enzymes or catalysts in a chemical
reaction, where the enzymes transform substrates in a continuous
manner without destroying themselves. The constant immune cell
population is assumed to act in a similar way, binding the tumor
cells and subsequently releasing them unable to replicate.

This work is devoted to the study of the Michaelis-Menten kinetics
using a stochastic approach. In natural biochemical systems, the
enzymatic activity of proteins depends on their configuration,
which may be very sensitive to the environment. Random
fluctuations of the environmental conditions (e.g. pH, ionic
strength) cause changes in enzyme activity. As a result of
this, the rate of product formation in the biochemical reaction
may deviate significantly from the mean. Because of the
exceptionally precise, accurate and efficient nature of biological
systems, these deviations may play a crucial role in the
functioning of the whole biological
system~\cite{Goel_Richter-Dyn}. On the other hand, fluctuations
may also play an important role in cancer growth. In the tumor
tissue, the growth rate and cytotoxic parameters are influenced by
many environmental factors, e.g. the degree of vascularization of
tissues, the supply of oxygen, the supply of nutrients, the
immunological state of the host, chemical agents, temperature,
radiations,  gene expression, protein synthesis and antigen
shedding from the cell surface, etc. As a result of this
complexity, it is unavoidable that in the course of time the
parameters of the system undergo random variations which give them
a stochastic
character~\cite{Lefever_Horsthemke,Garay,Goel_Richter-Dyn,Sica,Mantovani,Elliott}.

We will focus here on the interpretation of the Michaelis-Menten
reaction as a model for the process of tumor growth. The system is
subjected to fluctuating environmental conditions as a whole, and
to random variations of the kinetic parameter determining the
efficiency of immune response. Modelling the behavior of a tumor,
we will be most interested in the possibility of its spontaneous
extinction under the influence of random environmental
perturbations. In a certain range of  values of the parameters
describing the immune response intensity and the maximum density
of tumor, the model possesses two stable states: the state of
extinction, where no tumor cells are present, and the state of
stable tumor, where its density does not increase but stays at a
certain constant level. Random fluctuations present in the system
can induce transitions between those two states. From this point
of view, it is interesting to study the mean time of transition
from the tumor of a given density (e.g., from the state of a
stable tumor) to the state of extinction~\cite{anna}. We want to
find out how the extinction time can be changed by varying the
noise parameters, such as intensity or correlation time.

We report the possibility of onset of two noise-induced resonant
phenomena in the stochastic Michaelis-Menten reaction under study:
the resonant activation and the noise-enhanced stability. In the
model of tumor growth, these effects are interpreted as a
significant acceleration, or, respectively, deceleration of the
spontaneous extinction of tumor. Moreover, we show that the effect
of co-occurrence of both mentioned phenomena is possible: A region
can be found in the space of noise parameters, where
noise-enhanced stability is strongly reduced by resonant
activation. An important part of this work is the analysis of the
variance of first passage times compared to its mean as a function
of the noise intensity, which allowed us to explain in
detail the mechanisms underlying the studied phenomena.

The following section presents the model system used for the
analysis of cancer growth kinetics. In the next paragraph we report
the results and the appearance of RA and NES phenomena in this
model.

\section{The Model System}

\begin{figure}[t]
\epsfig{figure=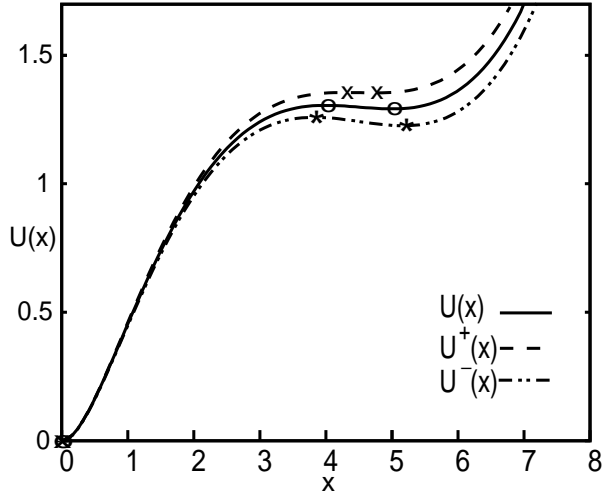, width=8cm} \caption{\label{fig:pot} The
Michaelis-Menten potential with parameters: $\beta=3, \theta=0.1,
\Delta=0.02$. Labels: ``o": extremes of $U(x)$:$x = 0, 4, 5$; ``x":
extremes of $U^+(x)$: $x=0, 4.28, 4.72$; ``*": extremes of $U^-(x)$:
$x=0, 3.83, 5.17$.
 }
\end{figure}

Biochemical reactions are usually described in terms of
phenomenological kinetic rates formulated by standard
stoichiometric analysis. In such deterministic models, molecular
fluctuations can be incorporated by including additional source of
stochastic fluxes  represented e.g. by an additive white noise
$\xi(t)$:
 \be
   \frac{dx}{dt}=f(x)-g(x)+\xi(t).
   \label{eq:lang}
 \ee
The above Langevin equation (further treated in the Ito
sense~\cite{kurtz-gardiner}) is based on a continuous description
of molecular species: time evolution of an input or output
concentrations $x$ produced at a rate $f(x)$ and degraded at rate
$g(x)$ defines a deterministic flux of reacting species and can be
used when modelling processes involve sufficient concentrations of
reacting agents. A Langevin equation with an additive driving
noise term is useful to describe molecular fluctuations in terms
of infinitesimal changes in an average mesoscopic variable $x$,
i.e. in concentrations. In that form Eq.~(\ref{eq:lang}) will be
further postulated to study dynamics of a catalytic reaction. A
``free-energy" profile $U(x)=-\int (f(x)-g(x)) dx$ for this
reaction is directly derived from the phenomenological
Michaelis-Menten scheme for the catalysis accompanying a
spontaneous replication of molecules:
 \bea
  X &\longrightarrow& \!\!\!\!\!\!\!\!\!\!^{\lambda} \quad 2X  \nonumber\\
 X\ +\ Y\  &\longrightarrow&
 \!\!\!\!\!\!\!\!\!\! ^{k_{1}}\ \ \ \ Z\ \longrightarrow
 \!\!\!\!\!\!\!\!\! ^{k_{2}}\ \ \ \  Y\ +\ P\ ,
 \eea
where a substrate $X$ forms first a complex $Z$ with molecules of
the enzyme $Y$, before the conversion of $X$ to a product $P$ is
completed. By assuming that the production of $X$-type molecules
inhibited by a hyperbolic activation is the slowest process under
consideration and by considering a conserved mass of enzymes
$Y+Z=E=const$, the resulting kinetics can be recast in the form of
the Langevin equation
\be
 \frac{dx}{dt}=-\frac{dU(x)}{dx}+\xi(t)
\ee
with the potential $U(x)$ expressed as
 \bea
  \label{eq:pot}
  U(x)=-\frac{x^2}{2}+\frac{\theta x^3}{3}+\beta x -\beta
  \ln(x+1),
  \label{prof}
 \eea
where $x$ is the normalized molecular density with respect to the
maximum number of molecules, and with the following scaling
relations
 \bea
 x=\frac{k_{1}x}{k_{2}}, \,\,\,\,\,   \theta=\frac{k_{2}}{k_{1}}, \,\,\,\,\,
\beta=\frac{k_{1}E}{\lambda}, \,\,\,\,\, t=\lambda t.
\label{eq:scaling}
 \eea
Note that the very same approach aimed to reduce the dimension of
the chemical master equation by employing so called
quasi-steady-state approximation~\cite{hanggi,samoilov-arkin}
(applicable when a subset of species is at steady state at the time
scale of interest) is broadly used in description of stochastic
chemical kinetics. In particular, for the enzymatic reactions
described by the Michaelis-Menten scheme, the quasi-steady-state
approximation assumes much larger concentration of substrates $X$
than the enzyme concentration $E$. This, by considering the
steady state constraint $dZ/dt=0$, becomes equivalent to assuming
that the propensity function for the decay of $X$ molecules is
given, in rescaled variables (Eq.~(\ref{eq:scaling})), by
$-\beta x/(1+x)$. The resulting potential profile Eq.~(\ref{prof})
has at most three extremes representing deterministic stationary
states of the system (see Fig.~\ref{fig:pot}):
\bea x_1&=&0, \\
 x_2&=&\frac{1-\theta+\sqrt{(1+\theta)^{2}-4\beta
   \theta}}{2\theta}, \\
 x_3&=&\frac{1-\theta-\sqrt{(1+\theta)^{2}-4\beta
   \theta}}{2\theta}.
 \eea

The essential feature captured by the model is, for a constant
parameter $\theta$, the $\beta$-dependent bistability. In the
above form and by assuming time dependent random variations of the
parameter $\beta$, the model has been used to describe  an effect
of cell-mediated immune surveillance  against the
cancer~\cite{Garay} or, alternatively, to analyze the
stochastic amplification and signaling in futile enzymatic
cycles~\cite{samoilov-arkin}. The immunological defense mechanisms
involve cell-mediated responses that consist in each case of
recognition processes of the cancer followed by their destruction
as proliferating cells. Most of tumoral cells bear antigens which
are recognized as strange by the immune system. A response against
these antigens may be mediated either by immune cells such as
T-lymphocytes or other cells, not directly related to the immune
system (like macrophages or natural killer cells). The process of
damage to tumor proceeds via infiltration of the latter by the
specialized cells, which subsequently develop a cytotoxic activity
against the cancer cell-population. The series of cytotoxic
reactions between the cytotoxic cells and the tumor tissue have
been documented
 to be well approximated~\cite{Garay} by a saturating,
enzymatic-like process whose time evolution equations are similar
to the standard Michaelis-Menten  kinetics. The variability of
kinetic parameters defining this process  naturally affects the
extinction of the tumor~\cite{Garay,Lefever_Horsthemke} and points
to the special role played by the parameters $\theta$ and $\beta$,
as described in~\cite{Lefever_Horsthemke}. The T-helper
lymphocytes and macrophages, can secrete cytokines in response to
stimuli. The functions that cytokines induce can both "\emph{turn
on}" and "\emph{turn off}" particular immune
responses~\cite{note}. The immune modulating activities of
cytokines are a result of their influence on gene expression,
protein syntesis and antigen shedding from the cell
surface~\cite{Sica, Mantovani, Elliott}. This "\emph{on-off}"
modulating regulatory role of the cytokines is here modelled
through a dichotomous random variation of the parameter $\beta$,
by taking into account the natural random fluctuations always
present in biological complex systems.


The mean escape time has been intensively studied in order to
characterize the lifetime of metastable  states of static and
fluctuating potentials with different initial
conditions~\cite{doe,bie,iwa,boguna,rei,bdka,ms,dybiec,man,as,dub,ps,ale}.
These studies show that the mean escape time has different
non-monotonic behaviors as a function of both the thermal noise
intensity and the mean frequency of potential fluctuations. These
behaviors are a signature of two noise-induced effects, namely the
resonant activation
(RA)~\cite{doe,bie,iwa,boguna,rei,bdka,ms,dybiec} and the noise
enhanced stability (NES)~\cite{man,as,dub,ps,ale}.
 NES, a phenomenon described theoretically and observed experimentally and
numerically in different physical systems, stabilizes a fluctuating
metastable state in such a way that the system remains in this state
for a longer time than in the absence of white noise. On the other
hand, due to the RA phenomenon, the mean escape time from the
metastable state across the fluctuating barrier may exhibit
non-monotonic dependence on the characteristic time scale of these
fluctuations.

In the model system investigated here we consider a chemical
Langevin equation with two independent sources of external
fluctuations represented by an additive driving noise term and a
dichotomous Markovian noise multiplying the $\beta$ parameter (see
Eq.~(\ref{eq:scaling})). The latter is responsible for regulatory
inhibition of the population growth and represents the modulating
stochastic activity of the cytokines on the immune
system~\cite{Sica, Mantovani, Elliott}. This process can change the
relative stability of metastable states and, in consequence, reverse
the direction of the kinetic process at hand. It should be
emphasized that the above mentioned phenomena of RA and NES act
counter to each other  in the cancer growth dynamics: the NES effect
increases in an unavoidable way the average lifetime of the
metastable state (associated to a
 fixed-size tumor state), while the RA phenomenon minimizes this lifetime.
Therefore, the purpose of this work is to find the optimal range of
parameters in which the positive role of resonant activation
phenomenon, with respect to the cancer extinction, prevails over the
negative role of NES, which enhance the stability of the tumoral
state.

We adhere to the model of  an overdamped Brownian particle moving
in a potential field between absorbing and reflecting boundaries
in the presence of noise which modulates the barrier  height. The
evolution of a state variable $x(t)$ is described in terms of the
Langevin equation
\begin{eqnarray}
\frac{dx}{dt} & =& -\frac{dV(x,t)}{dx}+\sigma\xi(t), \nonumber \\
  \nonumber \\
    V(x,t) &=& U(x) + G(x) \eta(t).
 \label{eq:lang2}
\end{eqnarray}

\noindent Here $\xi(t)$ is a Gaussian process with zero mean and
correlation function $\langle \xi(t)\xi(t')\rangle=\delta(t-t')$,
and $\sigma$ is the noise intensity. The potential
$V(x,t)$ is the sum of two terms: the fixed potential $U(x)$ and
the randomly switching term $G(x) \eta(t)$, where $\eta(t)$ stands
for a Markovian dichotomous noise switching between two levels
$\{\Delta^{+},\Delta^{-}\}$ with correlation time $\tau$ and mean
frequency $\nu = 1/(2 \tau)$. This means that its autocorrelation
function is
$$\langle(\eta(t)-\langle\eta\rangle)(\eta(t')-
\langle\eta\rangle)\rangle=\frac{\left(\Delta^+-\Delta^-
\right)^2}{4}e^{-|t-t'|/{\tau}}.$$

\noindent Both noises are assumed to be statistically independent,
\textit{i.e.} $\langle\xi(t)\eta(s)\rangle=0$. The potential
$V(x,t)$ therefore flips at random time between two configurations
\be U^{\pm}(x) = U(x) + G(x)\Delta^{\pm}.
\label{pot}
\ee

\noindent Based on Eq.~(\ref{eq:lang2}), we can write a set of
Fokker-Planck equations which describe the evolution of probability
density of finding the state variable
 in a ``position'' $x$ at time $t$:
\begin{eqnarray}
\partial_t {p}(x,\Delta^\pm,t)& =&  \partial_x  \left[\frac{dU^\pm(x)}
{dx}+\frac{1}{2}\sigma^2\partial_x  \right]p(x,\Delta^\pm,t) \nonumber \\
  & -& \frac{1}{2\tau} p(x,\Delta^\pm,t)+\frac{1}{2\tau} p(x,\Delta^\mp,t).
\label{eq:schmidr}
\end{eqnarray}

\noindent In the above equations time has dimension of
$[length]^2/energy$, due to a friction constant that has been
``absorbed'' in a time variable. With the initial condition
\be p(U\pm,x_s,t)|_{t=0}=\frac{1}{2}\delta(x-x_s), \ee

\noindent from Eqs.~(\ref{eq:schmidr}) we get the equations for the
mean first passage times (MFPTs):
\bea\label{eq:mfpt bn} -1 &=& -\frac{T^+(x)}{\tau}
+\frac{T^-(x)}{\tau}-
2\frac{dU^+(x)}{dx}\frac{dT^+(x)}{dx}
+\sigma^2\frac{d^2 T^+(x)}{dx^2}\nonumber \\
-1 &=& \frac{T^+(x)}{\tau}-\frac{T^-(x)}{\tau}-
2\frac{dU^-(x)}{dx}\frac{d T^-(x)}{dx}
+\sigma^2\frac{d^2 T^-(x)}{dx^2},
 \eea
where $T^+(x)$ and $T^-(x)$ denote MFPT for $U^+(x)$ and $U^-(x)$,
respectively. The overall MFPT  for the system reads
\bea
T(x)=T^+(x)+T^-(x)\label{eq:mfpt}, \eea
with boundary conditions
\bea \frac{dT^{\pm}(x)}{dx}|_{x=a}=0,\nonumber \\
T^{\pm}(x)|_{x=b}=0,
 \eea
which correspond to a reflecting boundary at $x=a$ and an
absorbing boundary at $x=b$. As in the usual physical picture of
resonant activation phenomenon \cite{doe} we expect that, for the
frequency of potential switching tending to zero (long correlation
time of the dichotomous noise $\eta(t)$), MFPT for the switching
barrier will be a mean value of MFPTs for both configurations
 \bea
\lim_{\tau\rightarrow\infty}T(U^+,U^-,\tau)
=\frac{1}{2}\left(T^+ + T^-\right), \eea
where the $T^+$ and $T^-$ are obtained for fixed
potentials with $U^+$ and $U^-$ respectively (Eq.~(\ref{pot})). For
the switching frequency tending to infinity (short correlation
time), the system will ``experience'' a mean barrier
\be \lim_{\tau\rightarrow 0}T(U^+,U^-,\tau)=
T\left(\frac{U^+}{2}+\frac{U^-}{2}\right). \label{eq:mfpt_infty}
\ee

Although the solution of the system~(\ref{eq:mfpt bn}) is usually
unique \cite{doe}, a closed, ``ready to use'' analytical formula
for MFPT can be obtained only for the simplest cases of
potentials. More complex cases require either use of approximation
schemes \cite{bie,rei,boguna}, perturbative approach \cite{iwa},
or direct numerical evaluation methods \cite{ms}.

The kinetics of our biological system is described by the equation
\be \frac{dx}{dt}=(1-\theta x)x - \beta\frac{x}{x+1},
\label{eq:system}
\ee
where $x(t)$ is the concentration of the cancer cells. The profile
of the corresponding quasi-potential (Eq.~(\ref{eq:pot})) presents a
double well with one of the minima at $ x=0 $. The region for $x>0$
can show either a monotonic behavior or a local minimum, depending
on the values of parameters $\theta$ and $\beta$. In the present
investigation we used only parameters able to give a local minimum
of the Michaelis-Menten potential for $x>0$: $\theta=0.1$ and
$\beta= 3$ (see Fig.~\ref{fig:pot}). For $x \rightarrow \infty$ the
potential shows a strong cubic repulsion. Taking into account a
random fluctuating environment (temperature, chemical
agents, radiation, etc..), we joined an additive noise term
$\sigma\xi(t)$ to the Eq.~(\ref{eq:system}). In order to describe
realistic fluctuations in immune
response~\cite{Sica,Mantovani,Elliott}, we added a
dichotomous Markovian noise $\eta(t)$, of amplitude $\Delta$ and
mean correlation time $\tau$, to the $\beta$ parameter. A
contribution of this kind implies that the effective potential
switches between two conformational states $U^\pm(x)$. Taking into
account all the noise contributions, we obtain the stochastic
Michaelis-Menten potential
 \be
U^\pm(x)=-\frac{x^2}{2}+\frac{\theta x^3}{3}+(\beta \pm \Delta)( x
- \ln(x+1)), \ee
and the Langevin equations for the system
 \bea
\dot{x} &=& -\frac{dU^\pm(x)}{dx} + \sigma \xi(t) \nonumber \\
&=& x(1-\theta x) - (\beta \pm \Delta) \frac{x}{x+1} + \sigma
\xi(t). \label{eq:langevin}
 \eea
The process of population growth and decay can be described as a
motion of a fictitious particle in the switching potential between
two configurations $U^+(x)$ and $U^-(x)$. The presence of noise
modulates the height of the barrier dividing two stable states of
the population. Transitions from one state to the other (here: from
a fixed-size tumor to a cancer-free state or \emph{vice versa}) are
induced by an  additive thermal-like noise~\cite{note2}. For
negligible additive noise and small concentration of tumor cells,
this model resembles a standard Verhulst equation with perturbing
multiplicative dichotomous noise, which exhibits a complex scenario
of noise-induced transitions, observable in a pattern of the
stationary probability density~\cite{Horsthemke_Lefever}. Here, we
will address kinetic properties of this model by studying the mean
first passage time (\ref{eq:mfpt}) between high and low population
states in the system. We will study how the two different sources of
noise as well as the position of the starting point
$x_{\mathrm{in}}$ influence the mean first passage time. We put the
absorbing boundary at $x=0$ and the reflecting one at $x=\infty$.
The event of passing through the absorbing boundary
 is equivalent to a total extinction of cancer.


\section{Results}

In order to obtain the MFPT for various starting points
$x_{\mathrm{in}}$, we performed a series of Monte-Carlo
simulations of the stochastic process (\ref{eq:langevin}) with an
absorbing boundary at $x=0$, reflecting boundary at $x=+\infty$
and the values of parameters: $\beta=3$, $\theta=0.1$, $\Delta=\pm
0.02$. The statistics for each MFPT has been taken from $10^3$
simulation runs, except for the results shown in
Figs.~\ref{fig:3d2}~-~\ref{fig:std_dev} where $10^4$
simulation runs were performed. The results confirm the existence
of resonant activation and noise-enhanced stability phenomena in
the studied system. Moreover, we have shown that in a certain
range of parameters both effects can occur together.

\begin{figure}[t]
\epsfig{figure=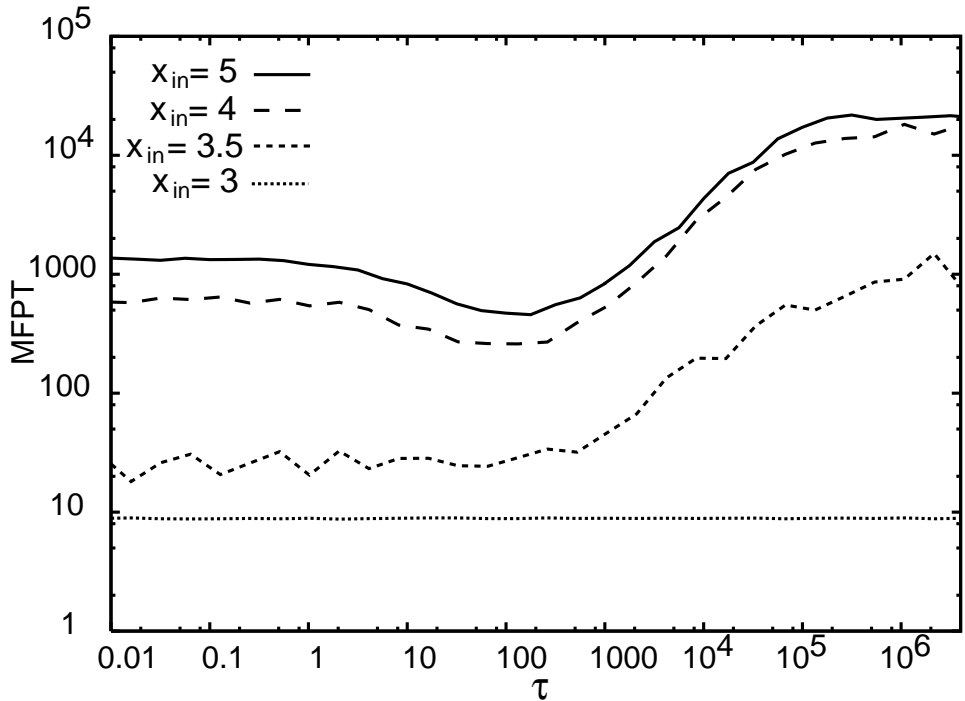, width=13cm} \caption{\label{fig:ra}
MFPT as a function of correlation time $\tau$ of the
dichotomous noise, for various starting points $x_{\mathrm{in}}$ in
the Michaelis-Menten potential. The RA effect is well
visible for initial positions near the metastable state and near the
maximum of the potential. Initial positions: $x_{\mathrm{in}}=5$
(in the neighborhood of the right minimum), $x_{\mathrm{in}}=4$ (in
the neighborhood of the maximum), $x_{\mathrm{in}}=3.5$ (left
slope),
 $x_{\mathrm{in}}=3$ (left slope). The parameter values are: $\beta = 3$,
 $\theta = 0.1$, $\Delta = 0.02$, and $\sigma = 0.1$.}
\end{figure}

\begin{figure*}[h]
\epsfig{figure=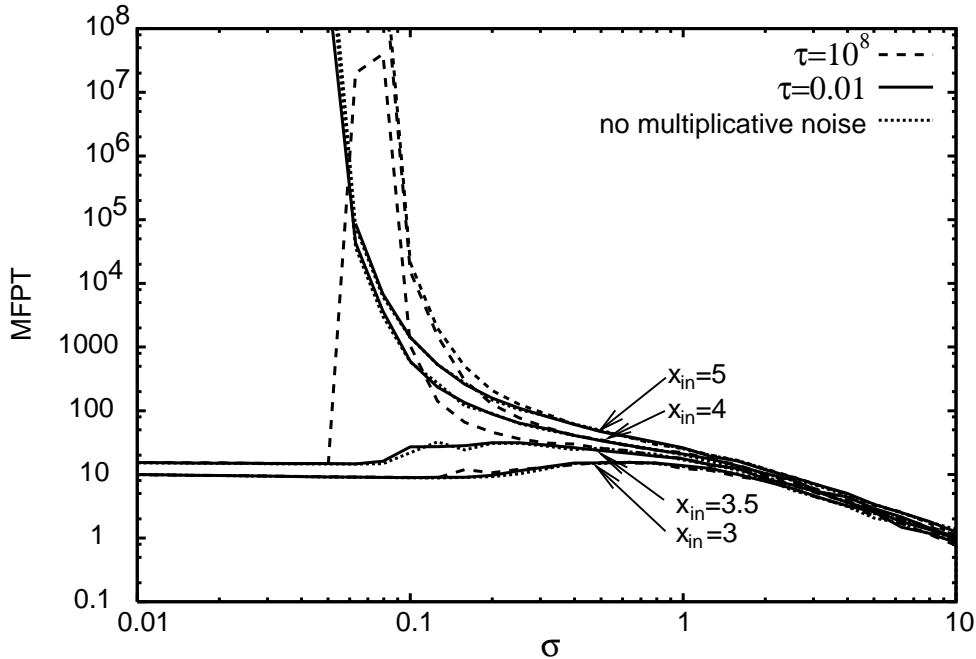, width=13cm} \caption{\label{fig:przekroj}
MFPT as a function of the noise intensity $\sigma$.
Influence of the multiplicative dichotomous noise on the NES effect
for the same initial positions as in Fig.~\ref{fig:ra} and two
values of correlation time of the dichotomous noise, namely: $\tau =
0.01$ (solid line) and $\tau = 10^8$ (dashed line). All the
other parameters are the same of Fig.~\ref{fig:ra}. The MFPTs in the
absence of multiplicative noise (dot lines), that is for an average
fixed barrier, are compared with those calculated at very low
correlation time $\tau$ (solid line).}
\end{figure*}
\begin{figure*}[h]
\epsfig{figure=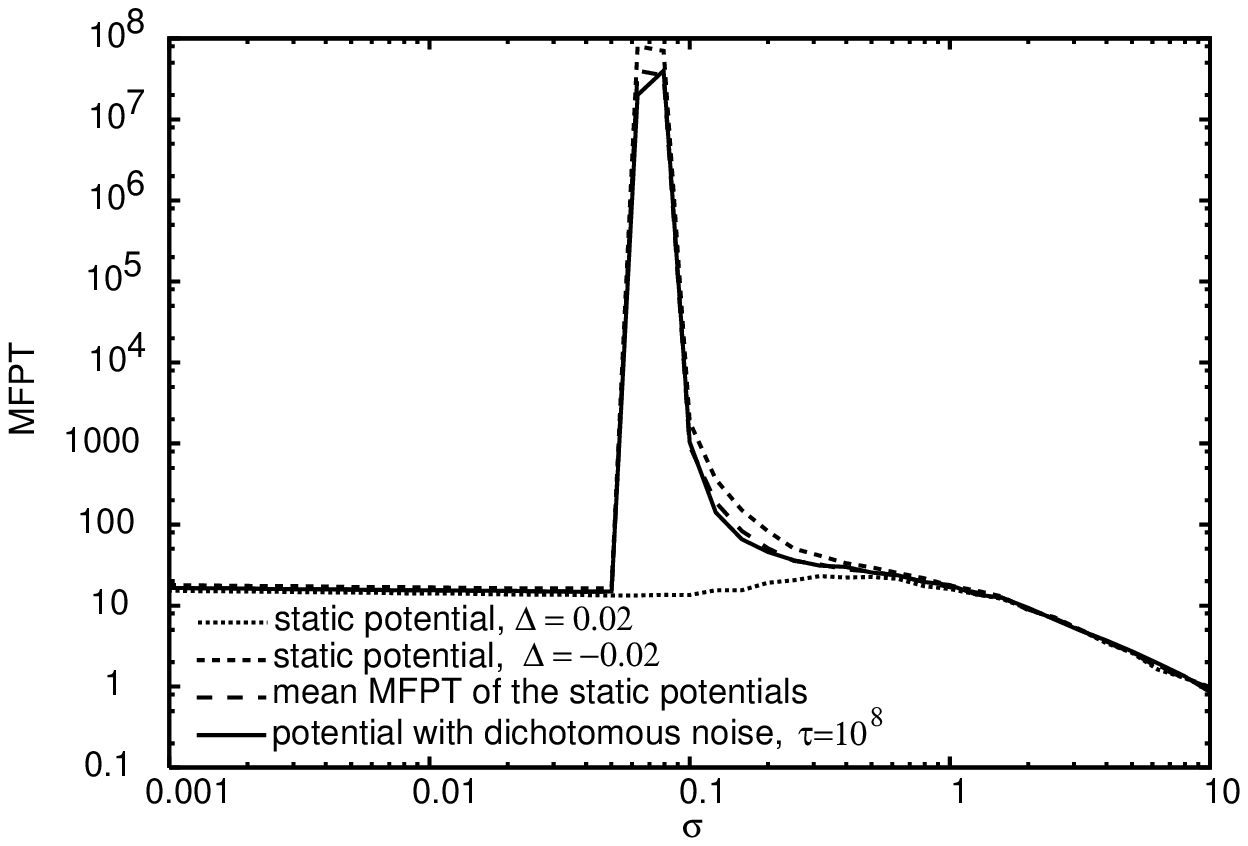, width=13cm} \caption{\label{fig:beta} MFPT
for the higher (dot line) and lower (short dashed line) position of
the switching potential and their mean, as a function of the
noise intensity $\sigma$. Although $\Delta$ is small, the NES
effect in a static potential with $\beta + \Delta$ looks quite
different from that obtained with $\beta - \Delta$. The mean MFPT of
the static potentials (long dashed line) is compared with the MFPT
calculated at very small switching frequency of the dichotomous
noise. The initial position is $x_{\mathrm{in}} = 3.5$, and all
the other parameter values are the same as in
Fig.~\ref{fig:ra}.}
\end{figure*}

\subsection{Resonant activation}\label{subsec:ra}

The resonant activation phenomenon occurs when, at a given
intensity $\sigma$ of the additive noise, there exists a certain
mean frequency $\nu_{\mathrm{min}}$ (and correspondingly a correlation time
$\tau_{\mathrm{min}}$) of the multiplicative noise,
 at which the mean first passage time is shortest.
 The resonant activation effect minimizes
 the average lifetime of a population in the metastable state.
Let us assume that the Brownian particle is behind the potential
barrier, in a neighborhood of the metastable state. When the barrier
fluctuations are very fast ($\tau$ small), the particle ``can never
adjust" to the instantaneous slope of the potential. Instead, it
``perceives" a slope which is an average of the higher and lower
configurations. The MFPT will tend to a constant value corresponding
to the average static potential. If the barrier fluctuations are
slower than the actual escape rate ($\tau$ large), the particle will
escape before any barrier flip occurs. Therefore, the mean first
passage time also tends to a constant, which now will be an average
of the escape times for the higher and lower configurations of the
potential. At the intermediate values of $\tau$, the escape rate is
an average of the escape rates for the higher and lower
configurations of the potential~\cite{bie,bdka}. In other words,
MFPT is the inverse of the mean escape rate, and, because of its
approximately exponential dependence on the ratio between the height
of the barrier and the noise intensity, its value is lower than both
above mentioned asymptotic mean first passage times~\cite{bdka}.

We compared the effect of resonant activation in the
Michaelis-Menten potential for trajectories starting from various
points: the neighborhood of the right minimum (bottom of the
potential well), the neighborhood of the barrier top, and two points
on the left slope of the barrier. In Fig.~\ref{fig:ra} we plot the
MFPT as a function of the correlation time $\tau$. The RA effect is
well visible for trajectories starting from behind the potential
barrier and even for those starting from its top. If the
trajectories start from the outer slope of the barrier, only a small
fraction of them can surmount it and be trapped in the potential
well, which would produce the effect of resonant activation. Most of
them rather run down the slope and approach the absorbing barrier
without having been trapped. This is the obvious reason why the
MFPTs for such initial points are shorter and their graphs less
accurate: The numerous contributions from the trajectories which
were not trapped are responsible for the shortening of MFPT. On the
other hand, only a small number of contributions from trapped
trajectories are responsible for the resonant activation effect.
Since the trapping events are very rare for $x_{\mathrm{in}}$ lying
on the outer slope, the statistics taken from a sample of $1000$
simulation runs turns out to be too small to produce a clear image
of RA.

\begin{figure}[h!]
\epsfig{figure=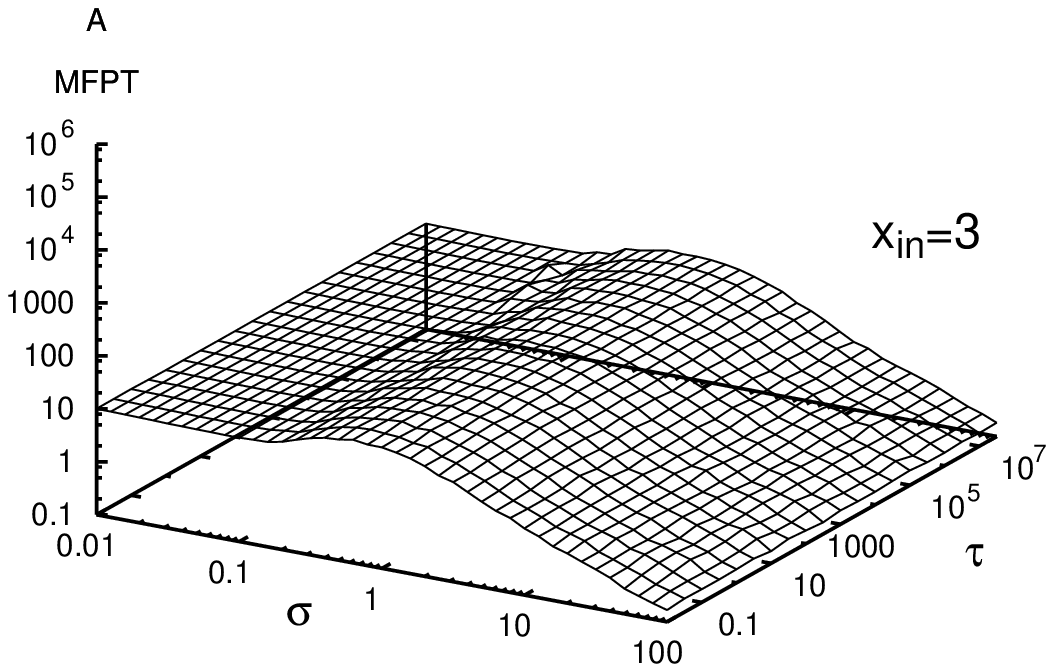, width=8cm}
\epsfig{figure=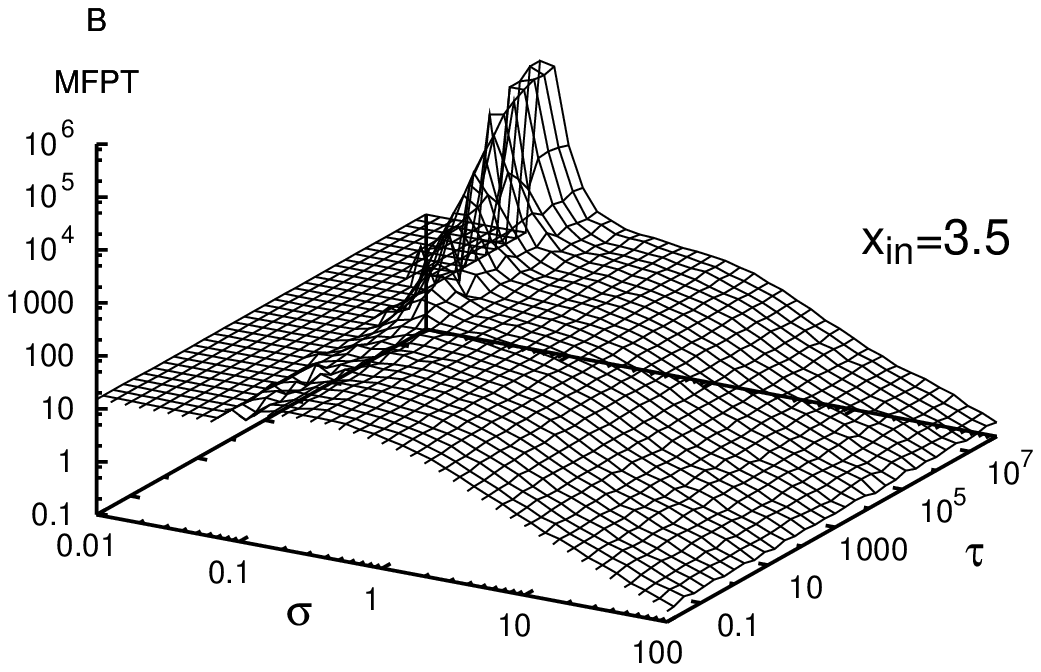, width=8cm}\\
\epsfig{figure=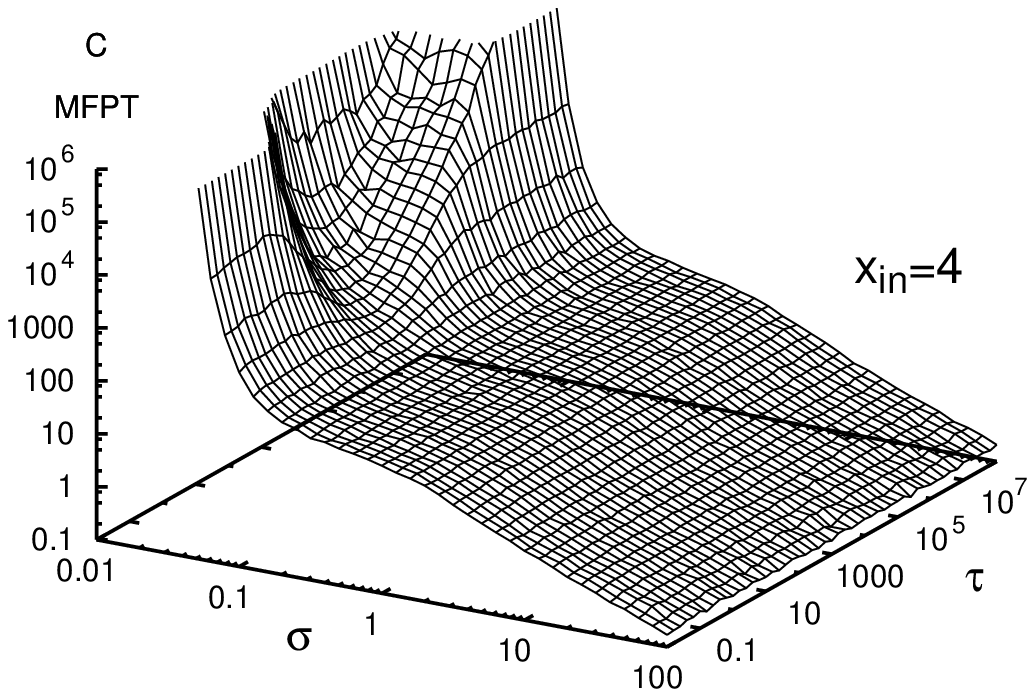, width=8cm} \caption{\label{fig:3d1} NES vs
RA: three-dimensional plot of MFPT as a function of noise
intensity $\sigma$ of the additive noise and correlation time $\tau$
of the dichotomous noise. The initial positions of the Brownian
particle and all the other parameter values are
the same as in Fig.~\ref{fig:ra}. }
\end{figure}
\begin{figure}[h!]
\epsfig{figure=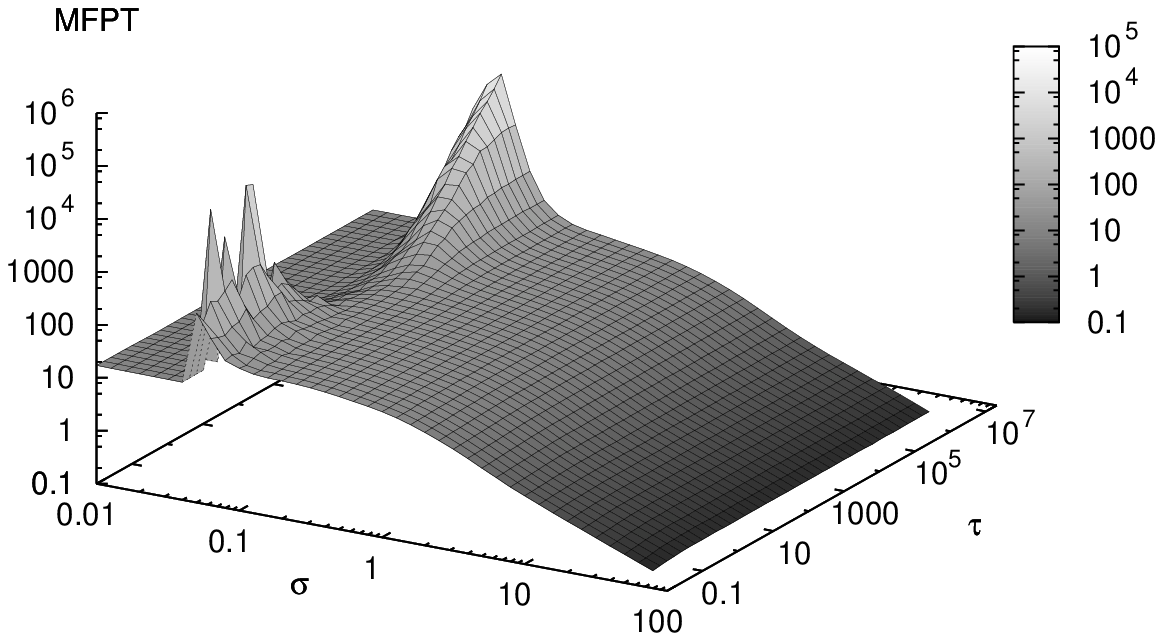, width=14cm} \caption{\label{fig:3d2} The
coexistence of NES and RA phenomena in a three-dimensional
plot of MFPT as a function of noise intensity $\sigma$ and
correlation time $\tau$. The initial position of the Brownian
particle is $x_{\mathrm{in}}=3.6$. Number of simulation
runs $N=10^4$. All the other parameter values are
the same as in Fig.~\ref{fig:ra}. }
\end{figure}

\subsection{Noise-enhanced stability}

The noise-enhanced stability effect occurs when, at a given mean
frequency $\nu$ of the multiplicative noise, there exists a certain
intensity $\sigma_{\mathrm{max}}$ of the additive noise, at which
the mean first passage time is longest. Differently from RA, the NES
effect maximizes the average lifetime of the population in a
metastable state as a function of the noise intensity. The
nonmonotonic behavior of the mean escape time as a function of the
additive noise intensity $\sigma$ depends on the potential profile
parameters, on the parameters of the multiplicative dichotomous
noise and also on the initial position of the Brownian
particle~\cite{as,dub,ale}. Noise enhances the stability of the
metastable state with different peculiarities related to different
dynamical regimes: the average lifetime can greatly increase when
the noise intensity is very low with respect to the height of the
barrier and the initial positions of the Brownian particles are in
the ``divergent" dynamical regime~\cite{ale}.

If the particle starts from initial positions within the potential
well, at small values of $\sigma$ it will rather stay trapped in the
well than escape from it, according to the Kramers formula. The mean
escape time will then tend to infinity for $\sigma \rightarrow 0$.
If the particle starts from outside the well, at small $\sigma$ it
will at once run down the slope and its mean escape time will tend
to the escape time of a deterministic particle. At high noise
intensities, the mean escape time decreases monotonically,
regardless of initial positions. At intermediate noise intensities,
a particle starting from the outer slope may sometimes be trapped
into the well. Such events, although rare, can significantly
increase the mean escape time because a trapped particle stays then
in the well for a relatively long time.

In Fig.~\ref{fig:przekroj} we show how the dichotomous barrier
switching influences the NES effect. If the mean switching frequency
$\nu$ is large, then the MFPT behaves as the mean first passage time
for an average barrier, so the NES effect looks like in a potential
with no multiplicative noise ($\Delta=0$). When, in turn, the mean
switching frequency $\nu$ is very small, then the MFPT is an average
of the mean first passage times for the higher and lower position of
the switching potential (see Fig.~\ref{fig:beta}). Here the NES
effect can sometimes differ considerably from the same effect in the
static potential, even if the amplitude of the dichotomous noise
$\Delta$ is very small with respect to the value of the $\beta$
parameter.

\subsection{Noise-enhanced stability vs. resonant activation}

In Figs.~\ref{fig:3d1} and~\ref{fig:3d2} we present the combined
view of RA and NES effects. We notice that RA is better visible when
the additive noise is weak compared to the height of the barrier.

If the potential barrier is high enough or the additive noise is
weak enough, the mean first passage time in a static potential can
be approximated by the inverse of the Kramers escape rate, which
increases exponentially with $\Delta U(x)/\sigma^2$, where $\Delta
U(x)$ is the height of the barrier. Such an exponential dependence
guarantees the existence of a minimum of MFPT$(\tau)$. Its value
at intermediate $\tau$ is lower than both asymptotic mean first
passage times, for very large or very small $\tau$. Specifically
the MFPT at very low mean switching frequency
$T_{\tau\rightarrow\infty}$ ($=
\frac{1}{2}\left[T^+ + T^-\right]$), will be higher than
the MFPT at the high frequency limit $T_{\tau\rightarrow 0}$
($=T\left(\frac{U^+}{2}+\frac{U^-}{2}\right)$). In the middle
frequency regime the effective escape rate over the fluctuating
barrier is the average of the escape rates ($\frac{1}{2}(K_+ +
K_-),$ where $K_+ = 1/T^+$ and $K_- = 1/T^-$,
with $K_+\gg K_-$, and because of the exponential
dependence, the MFPT in this intermediate frequency regime will be
smaller than $T_{\tau\rightarrow 0}$. The resonant activation
phenomenon therefore will manifestly occur if $\sigma \ll
\Delta(U^+_{\mathrm{max}} - U^-_{\mathrm{max}})$, where
$\Delta(U^+_{\mathrm{max}} - U^-_{\mathrm{max}})$ is the
difference in height between the higher and lower position of the
barrier at the maximum of the potential well~\cite{bdka}. In the
case of our potential, that difference is of order $10^{-1}$, so
the resonant activation can be observed for $\sigma < 10^{-1}$ as
we can see from Fig.~\ref{fig:3d1}.

To show better the RA phenomenon in the parameter region of NES,
we must choose an appropriate initial condition $x_{\mathrm{in}}$.
In order to obtain a well-visible RA (a large enough number of
trajectories trapped in the potential well) we have to choose the
initial position sufficiently close to the top of $U^-(x)$. We
report in the following Fig.~\ref{fig:3d2} the co-occurrence of
resonant activation and noise-enhanced stability at
$x_{\mathrm{in}}=3.6$. Since both effects act here in an opposite
way, there exists a regime of $\sigma$ and $\tau$ parameters where
noise-enhanced stability is strongly reduced by resonant
activation.

\begin{figure}[t]
\epsfig{figure=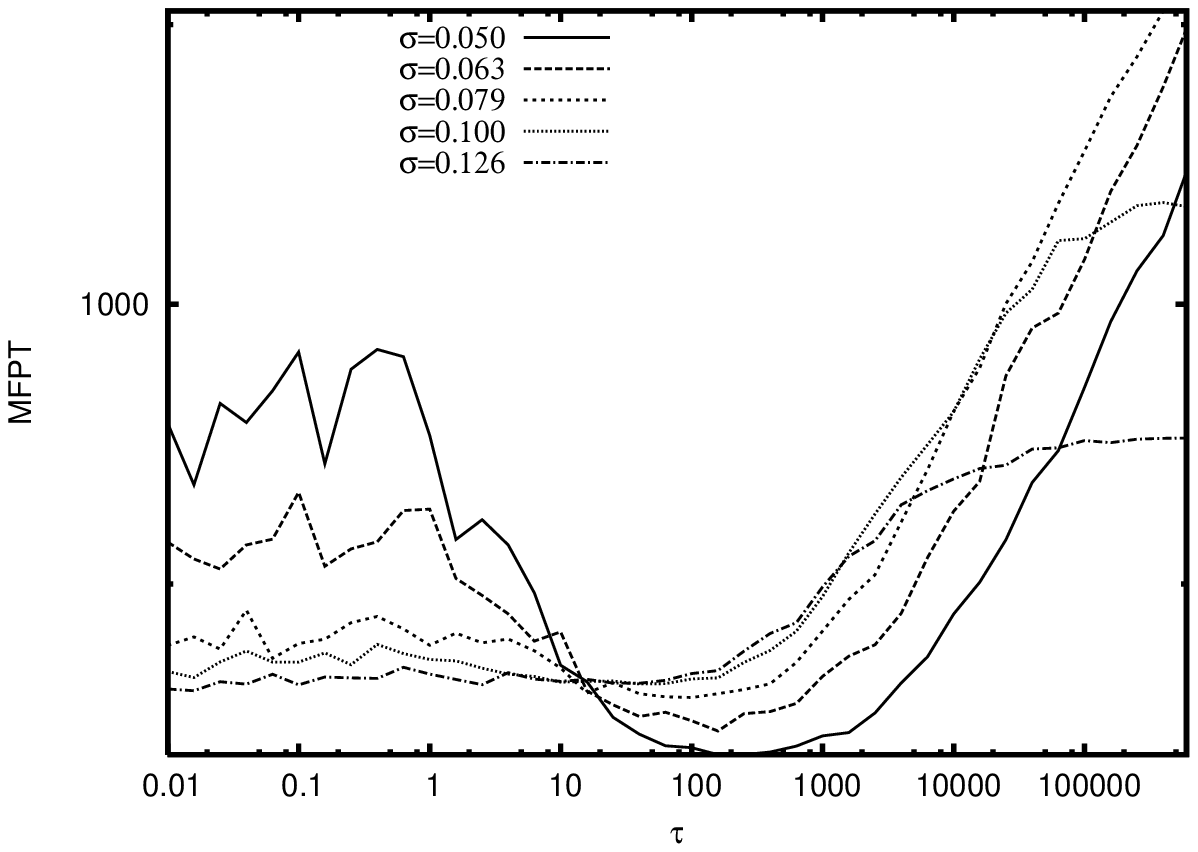, width=13cm} \caption{\label{fig:ra_3.6}
Cross-sections of Fig.~\ref{fig:3d2} for six values of the
noise intensity $\sigma$, in the region of coexistence of RA and
NES. The effect of resonant activation overlapping the
noise-enhanced stability is well visible. The initial point
is $x_{\mathrm{in}} = 3.6$. Number of simulation runs
$N=10^4$. All the other parameter values are the same as
in Fig.~\ref{fig:ra}.}
\end{figure}

\begin{figure}[t]
\epsfig{figure=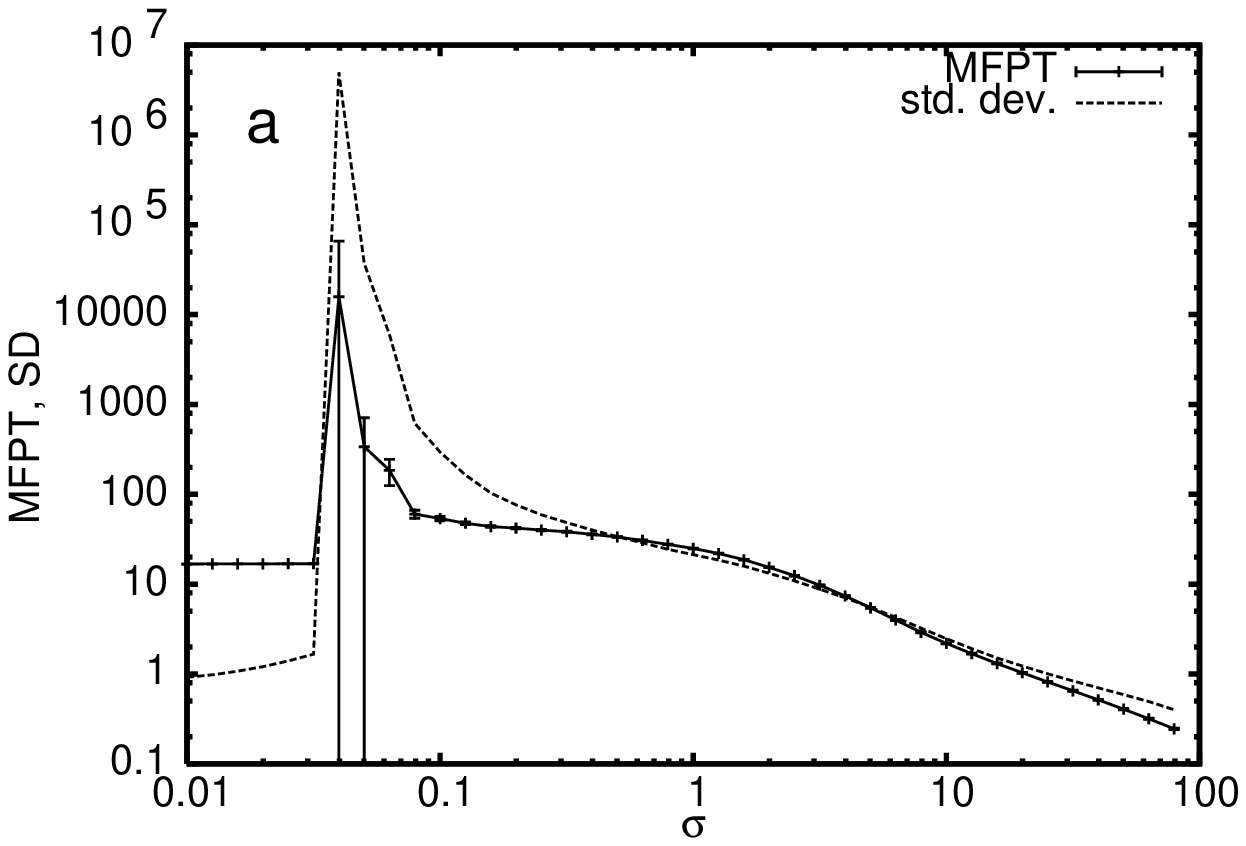, width=8cm} \epsfig{figure=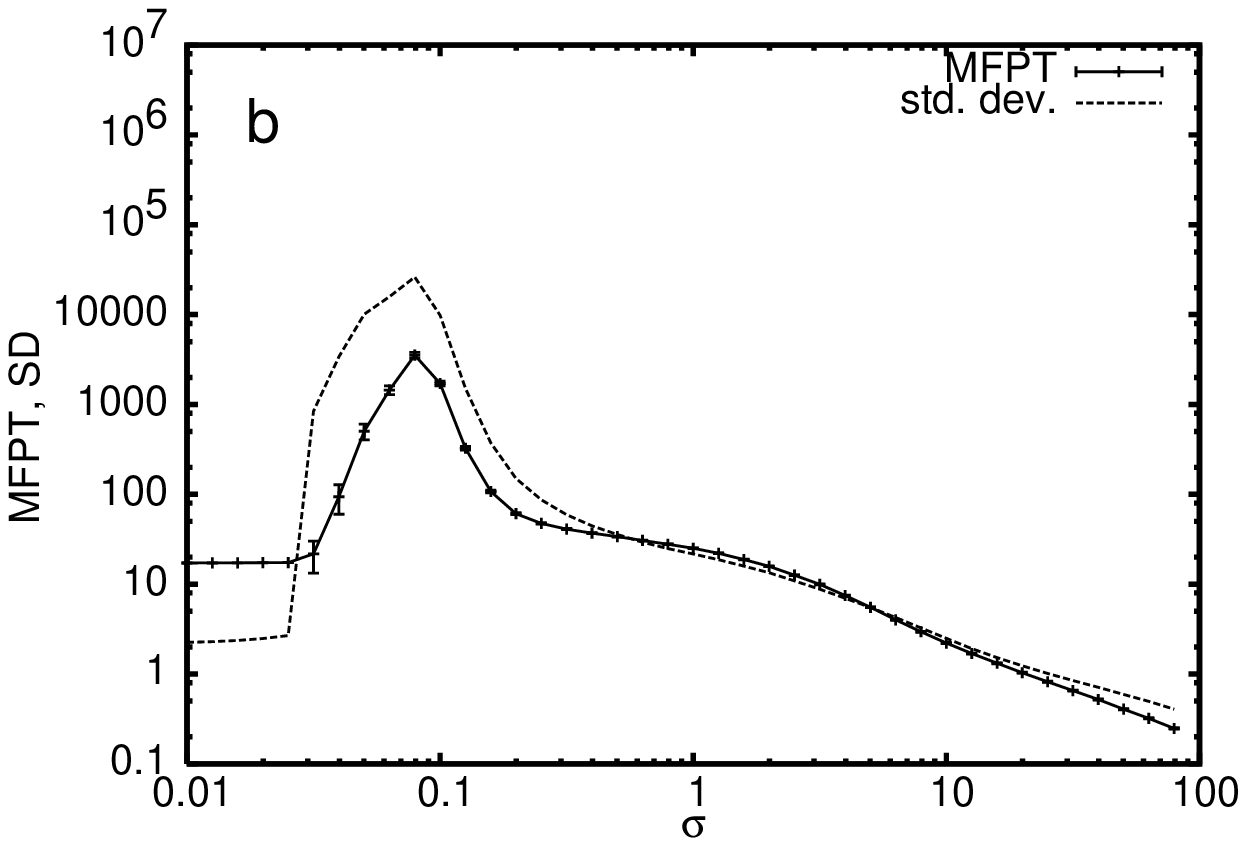,
width=8cm} \caption{\label{fig:std_dev} MFPT and its
standard deviation, with error bars, as a function of the additive
noise intensity $\sigma$ for two values of the correlation time
$\tau$, namely (a) $\tau = 1$ and (b) $\tau = 10^5$.  Number of
simulation runs $N=10^4$, initial position of the Brownian
particle $x_{in} = 3.6$. All the other parameter values are the
same as in Fig.~\ref{fig:ra}. The asymmetry of appearance of error
bars near the maximum in Fig.~\ref{fig:std_dev}a is due to the
large values of error bar in these points and to the log scale.}
\end{figure}

The above considerations are valid not only for trajectories
starting from inside the potential well, but also for arbitrary
initial positions in the potential, if only the particle has a
chance to be trapped behind the potential barrier for some time.
If the additive noise intensity is very large, a particle starting
from the outer slope of the barrier does not ``feel" the barrier
at all. If, in turn, the additive noise is very weak, the particle
slides down the slope in an almost deterministic way. But at
intermediate values of $\sigma$, the particle can (in some
realizations) be trapped behind the barrier and, in case of such a
rare event, its escape time changes non-monotonically as a
function of $\tau$, in a way described in
subsection~\ref{subsec:ra}. This effect of coexistence of
noise-enhanced stability and resonant activation effects can be
observed in Fig.~\ref{fig:3d2}. In Fig.~\ref{fig:ra_3.6}
we report the cross-sections of Fig.~\ref{fig:ra}, in the region
of coexistence of RA and NES, for five values of the additive
noise intensity $\sigma$, namely: $\sigma = 0.050, 0.063, 0.079,
0.100, 0.126$. The effect of the RA overlapping the NES is well
visible. It is clear that for noise intensity values greater than
$\sigma\approx 0.1$, the nonmonotonic behavior typical of RA
effect disappears.

In Fig.~\ref{fig:std_dev} we present the graphs of MFPT vs. additive
noise intensity $\sigma$ together with the standard deviation of
$N=10^4$ first passage times, at $\tau=10^5$ (slow
switching) and $\tau=10^0$ (fast switching). The initial
position has been set up to $x_{\mathrm{in}}=3.6$. At large
intensities of the additive noise the standard deviation is equal to
MFPT, which indicates the purely Arrhenius behavior of the system
kinetics~\cite{note3}. This can be explained in the following way:
if the additive noise is strong enough, the Brownian particle does
not "feel" the random changes in the height of potential barrier and
its motion becomes  diffusive. In the region of NES, the system's
behavior begins to diverge from the Arrhenius model and typically,
the standard deviation is much higher than the mean of first passage
times. On the other hand, at small intensities of the additive
noise, the standard deviation is much lower than the mean. In the
range of $\sigma$ where the NES occurs, the possible trajectories
can be roughly divided into the "short" ones, which were not trapped
but  ran down to the absorbing boundary at $x=0$, and the "long"
ones, which were trapped behind the potential barrier for a long
time. In a sample of registered $N=10^4$ first passage times, we
observe  a large number of "short" times of order of $10$, whose
standard deviation is of order of $1$, and a small number of "long"
times of order of $10^5$ (at $\tau=10^5$) or $10^7$ (at $\tau=10^0$)
whose standard deviation is of order of $10^5$ (at $\tau=10^5$) or,
respectively, $10^7$ (at $\tau=10^0$). The contribution of "long"
trajectories to the overall standard deviation makes it very large.
However, the weaker the additive noise is, the smaller is the number
of trapped trajectories and their contribution to the standard
deviation. In the region of small $\sigma$, where the standard
deviation is smaller than MFPT, no trapped trajectories were
recorded. It suggests that the region is dominated by an almost
deterministic kinetics with contributions from the stochastic
diffusion visible mostly in a flat part of the potential well close
to $x=0$.

It should be noted, however, that the behavior of MFPT
shown in Fig.~\ref{fig:std_dev}a, at the fast switching of the
potential profile ($\tau = 1$) and with the initial position
$x_{in} = 3.6$, corresponds, in the limit of $\tau \rightarrow 0$,
to a Brownian particle moving in a fixed potential (the average
potential profile $U(x)$ of Fig.~\ref{fig:pot}) starting from an
initial unstable state and in a divergent dynamical regime with
regards to the MFPT (see Ref.~\cite{ale} for a detailed discussion
on this point). In this regime both the MFPT and its standard
deviation (SD) diverge for $\sigma \rightarrow 0$. Because of the
finiteness of the ensemble of realizations and of the observation
time considered in our numerical experiments we don't observe the
divergence of MFPT and its SD, and for $\sigma \rightarrow 0$ the
deterministic escape time is obtained. While the behavior shown in
Fig.~\ref{fig:std_dev}b corresponds to the "\emph{real}"
nonmonotonic behavior of MFPT and its SD as a function of the
noise intensity $\sigma$. These different dynamical regimes
experienced by the Brownian particle explain the two behaviors of
Fig.~\ref{fig:std_dev}. As a consequence we expect very large
error bars just in the maximum value of MFPT
(Fig.~\ref{fig:std_dev}a), and decreasing error bars by increasing
noise intensity values. This error bar behavior should be
considered as a signature of the divergent dynamical regime in the
NES effect.

%

\section{Conclusions}

We studied a Langevin equation derived from the phenomenological
Michaelis-Menten scheme for catalysis accompanying a spontaneous
replication of molecules. It contains an additive noise term
(Gaussian white noise) and a multiplicative noisy driving
(dichotomous noise) in the term responsible for inhibition of
population growth. This model may be used e.g. to describe an effect
of cell-mediated immune surveillance against cancer. Specifically
the relative rate of neoplastic cell destruction is a random
dichotomous process due to the action of cytokines on the immune
system.

 We examined how the two different sources of noise influence the
population's extinction time, identified with the mean first
passage time of the system over the zero population state.

The first result obtained concerns the confirmation of the existence
of noise enhanced stability (NES) phenomenon in a system in the
absence of multiplicative noise as well as in its presence. The NES
effect increases the extinction time as a function of the intensity
of additive noise. The second result we have found in our study is
the presence of resonant activation (RA) effect. This phenomenon
minimizes the extinction time as a function of the correlation time
of multiplicative noise. The two effects are acting in an opposite
way in the cancer growth dynamics. Namely, the NES effect by
increasing the lifetime of the metastable state delays the escape
from the tumor state, while the RA effect by decreasing this
lifetime leads to the extinction of the tumor. An appropriate choice
of values of the noise parameters allows either to maximize or
minimize the extinction time of the population.

Another result is the evidence for the possibility of co-occurrence
of both mentioned noise-induced effects: resonant activation can be
observed also in the region of noise-enhanced stability. In this
coexistence region the NES effect which enhances the stability of
tumoral state becomes strongly reduced by the RA mechanism which
enhances the cancer extinction. In other words, an asymptotic
regression to the zero  tumor size may be induced by controlling the
noise affecting hyperbolic inhibition of a spontaneous proliferation
of cells.

An important part of this work is the analysis of the variance of
first passage times compared to its mean, as the function of the
noise intensity. This allowed us to fully explain the mechanisms
underlying the resonant phenomena observed: the Arrhenius diffusive
behavior in the range of high intensities of the additive noise; the
divergence from the latter due to "trapping" events in the
intermediate region where noise-enhanced stability occurs; and,
finally, the domination of deterministic kinetics, with
contributions from the stochastic diffusion visible mostly in a flat
part of the potential well.

Our simple theoretical model can help to obtain new insight into
the complexity of tumor progression, and at the same time may lead
to new experiments in which the modulating activity of cytokines
can be driven by chemical agents or light irradiation. Of course
there are many other factors characterizing the dynamics of the
cancer growth due to the complexity of the system and due to many
different types of cancer diseases.

\begin{acknowledgments}
This work was supported by the ESF funds (via the STOCHDYN program)
and MIUR. Additionally, A.O-M and E.G-N acknowledge financial
support from the Polish State Committee for Scientific Research
through the grants 1P03B15929 (2005-2007) and 2P03B08225.

We greatly thank  Dr. C. Tripodo, of the Inst. of Anatomy
and Pathological Histology of Palermo University, and Dr. A. Sica,
of the Inst. of Pharmacological Researches M. Negri
(Milano), for useful and deep discussions on the medical aspects of
the cancer growth dynamics.
\end{acknowledgments}

%

%


\begin{thebibliography}{100}
\bibitem{note2} The correspondence with a thermal noise is a loose one,
  since a process of the population growth and extinction goes on rather in a
 changing environment and the spontaneous fluctuations of the
 densities of cells do not follow the fluctuation-dissipation
 theorem.
\bibitem{note3}Exponential distribution of first passage times
describes the distance between events with uniform distribution in
time. If $t$ stands for time variable and $\lambda t$ is the
expected number of events in the interval $[0,t]$ then $e^{-\lambda
t}$ is the probability of no events in $[0,t]$ . Consequently, for
the exponential pdf $\lambda e^{-\lambda t}$  the ratio between the
first moment $<t>$ and the standard deviation $\sqrt{Var[t]}$ is
equal to 1.
\end{thebibliography}
\end{document}